\documentstyle[12pt]{article}
\newcommand{\be}{\begin{equation}}
\newcommand{\ee}{\end{equation}}
\newcommand{\bea}{\begin{eqnarray}}
\newcommand{\eea}{\end{eqnarray}}
\newcommand{\ba}{\begin{array}}
\newcommand{\ea}{\end{array}}
\newcommand{\beas}{\begin{eqnarray*}}
\newcommand{\eeas}{\end{eqnarray*}}
\newcommand{\bes}{\begin{equation*}}
\newcommand{\ees}{\end{equation*}}

\def\i2           {\mbox{$\frac{i}{2}$}}

\begin{document}

\title{\bf Klein Gordon particle near R-N  black  hole, generalized sl(2) algebra  and harmonic oscillator energy }

\author{J. Sadeghi $^{a}$\thanks{Email:pouriya@ipm.ir}\hspace{1mm} and  M. R.  Alipour $^{a}$\thanks{Email:m.r.alipour@cert.umz.ac.ir}\\
$^a$ {\small {\em  Sciences Faculty, Department of Physics, University of Mazandaran,}}\\
{\small {\em P.O.Box 47415-416, Babolsar, Iran}}}

\maketitle
\begin{abstract}
\noindent

In this paper, we  consider Klein Gordon  particle near  Reissner–Nordstr\"om   black hole. The symmetry of such background lead us to  compare the corresponding Laplace equation with the generalized Heun functions. Such relation help us achieve the generalized $sl(2)$ algebra and some suitable results for  describing the above mentioned symmetry . On the other hand in case of $r\rightarrow r_+$ which  is near the proximity black hole, we obtain the energy spectrum and wave function.  When  we compare  the equation of $R-N$ background with  Laguerre differential equation, we show that the obtained energy spectrum is same as  three dimensional harmonic oscillator.  So, finally we take advantage of  harmonic oscillator energy and make suitable partition function.  Such function help us to obtain all thermodynamical properties of black hole. Also, the structure of obtained entropy lead us to have some bit and information theory in the $R-N$ black hole .

{\bf Keywords:} R-N black hole; Three dimensional harmonic oscillator; Heun equation; generalized  sl(2) algebra

\end{abstract}
\noindent

\section{Introduction}
A charged black hole is interesting object which  possesses electric charge.  Here we note that, If we compare two force as a electrical and gravitational force to each other , we will see that the electrical charge is grater than to gravitational (by about 40 orders of magnitude). Or in comparing with the gravitational attraction an electrically charged mass is dramatically greater than the gravitational attraction. It is not expected that black holes with a significant electric charge will be formed in nature.
On the other hand, a charged black hole is one of three possible types of black holes that could exist in Einstein's theory of gravitation, general relativity.

The theoretical studies concerning physical processes which occur in the
spacetime surrounding black holes are of special interest. It can certainly  help
us to understand the physics of these interesting objects which is predicted by general
relativity [1]. As we know Reissner and Nordstrom found a similar solution as Schwarzschild, but it was  more complicated  with two parameters mass $M$ and charge $Q$. On the other hand, we know that if we assume some free particle move in corresponding $R-N$ background we will face  Klein Gordon  particle near Reissner–Nordström black hole. In order to investigate some symmetry during process we have to introduce some suitable function. So, during last years, the appropriate function was  Heun it,s generalization. These functions have gaining more and more importance due to the large number of applications in different areas of physics  and in special, in the solutions of problems related
to a scalar field in gravitational backgrounds [2]. Also,  the study of field equations in curved space-time is of interest for physical
considerations. The effect of space-time curvature on the behavior of micro
systems gives information in a general as well in a strong gravitational field.
The knowledge of the solution of field equations makes possible the evaluation
of the physical effects of the gravitational field [3]. We  use  the Heun’s functions
and  find the  solutions of the Klein-Gordon
equation in some black hole spacetimes [4]. Also, we take advantage from Heun equation and obtain  the $sl(2)$ algebra from the black hole background. Without use of
these functions, the scalar solutions for massless and massive scalar fields will
 be possible only for some specific regions very close and far away from the black
hole horizons. So the Heun equation  help us to modify equation in terms of first order operator which is shown the $sl(2)$ generators algebra. All above information give us motivation to arrange paper as follow. In section 2 we introduce Reissner–Nordström black hole and  write the Klein Gordon equation. This equation lead us to arrive at the corresponding Hamiltonian for the pointed black hole. We compare the above mentioned  results with the Heun equation and achieved  some raising and lowering operators which are form of generators and are satisfied by the generalized combative $sl(2)$ algebra. In section 3 we employ the particle on Reissner–Nordström black hole and obtain the energy and wave function in near Reissner–Nordström black hole.  By comparing the corresponding equation with Laguerre equation we obtain the energy spectrum.  We compare  the obtained energy spectrum   with harmonic oscillator energy. In section 4 we arrive at thermal properties of Reissner - Nordstrm black hole with energy spectrum. And also bit and entropy as a information subject are discussed by this section. Finally in section 5 we have some conclusion and also suggestion for the future research.

 \section{Reissner–Nordström black hole}
We consider a Reissner-Nordstrom black-hole spacetime of mass $M$ and electric charge $Q$ which is characterized by
the following curved line element, [5]
\begin{equation}
    ds^2=f(r)dt^2-\frac{1}{f(r)} dr^2- r^2 d\theta^2-r^2 sin^2\theta d\phi^2
\end{equation}

where $f(r)$ for the Reissner–Nordström  black hole will be as,
\begin{equation}
f(r)=1-\frac{r_s}{r}+\frac{r_Q^2}{r^2} \hspace{0.5cm}  r_Q=\frac{\sqrt{KG}Q}{c^2}    \hspace{0.5cm}   r_s=\frac{2GM}{c^2}
 \end{equation}
In here, we are going to write the Klein Gordon equation  which is given by,
\begin{equation}
  E^2=p^2c^2+E_0^2, \qquad     \Longrightarrow  \qquad       \hat{H}^2=\hat{P}^2c^2+\hat{{H_0}^2}.
\end{equation}
For the momentum operator $\hat{p}=\frac{\hbar}{i}\vec{\nabla_{bel}}$ the above equation becomes
 \begin{equation}
\hat{H}^2=-{\hbar}c^2{\nabla^2_{bel}}+\hat{{H_0}^2}
\end{equation}
In order to write the first terms of equation (4), one can consider the following Laplace–Beltrami operator,
 \begin{equation}
\nabla^2_{bel}=\frac{1}{\sqrt{g}}\partial_i(\sqrt{g}g^{ij}\partial_j)
\end{equation}
where $g_{ij}$ can be obtained by the corresponding black hole as,

\begin{equation}
 g_{ij}=\left(
\begin{array}{ccc}
 -\frac{1}{\emph{f(r)}} & 0 & 0 \\
 0 & -r^2 & 0 \\
 0 & 0 & -r^2sin^2\theta \\
 \end{array}
  \right)
  \end{equation}
By using the equations (5) and (6) into equation (4) we will arrive at following relation,
 \begin{equation}
\hat{H}^2={\hbar}c^2\left(\sqrt{\emph{-f(r)}}\partial_r(r^2\sqrt{\emph{-f(r)}}\partial_r)-\frac{\hat{L^2}}{\hbar^2r^2}+\frac{\hat{{H_0}}^2}{\hbar^2 c^2}\right)
\end{equation}
Now we choose the function as $\Psi(r,\theta,\varphi)=Y_\ell^m(\theta,\varphi) F(r)$, so we have following relations,
\begin{equation}
\hat{L^2} Y_\ell^m(\theta,\varphi)=\ell(\ell+1) Y_\ell^m(\theta,\varphi)\hspace{1.5cm}
\hat{{H_0}^2}\Psi={E_0}^2\Psi
\end{equation}

Finally, the Hamilton corresponding to Reissner–Nordström  black hole will be as,
\begin{equation}
\hat{H}^2={\hbar}c^2\left(\sqrt{\emph{-f(r)}}\partial_r(r^2\sqrt{\emph{-f(r)}}\partial_r)-\frac{\ell(\ell+1)}{\hbar^2r^2}+\frac{{E_0}^2}{\hbar^2 c^2}\right)
\end{equation}

Before going to solve the above equation, we simplify the $f(r)$ with two horizons for the Reissner–Nordström black hole as $r_-$ and  $r_+$ which are given by following,
\begin{equation}
\emph{f(r)}=1-\frac{r_s}{r}+\frac{r_Q^2}{r^2} =\frac{(r-r_-)(r-r_+)}{r^2}=0
\end{equation}
and
\begin{equation}
 r_-=\frac{r_s-\sqrt{{{r_s}^2}-{4r_Q}^2}}{2}     \hspace{2cm} r_+=\frac{r_s+\sqrt{{{r_s}^2}-{4r_Q}^2}}{2}
\end{equation}

Here, we consider  free particle as $(E_0=0)$ and Hamiltonian of system with the above mentioned ${f(r)}$ can be written by following relation,
\begin{eqnarray}
\hat{H^2}&=&\hbar c^2\frac{(r-r_-)(r-r_+)}{r^2}\nonumber\\&&\left[\partial_r^2+(\frac{1}{r}+\frac{1}{2(r-r_-)}+\frac{1}{2(r-r_+)})\partial_r-\frac{\ell(\ell+1)}{(r-r_-)(r-r_+)}\right]\nonumber\\
\end{eqnarray}

In order to have solution and algebraic discursion , we introduce the following Heun equation [6,7]
\begin{equation}
{F^{\prime\prime}}(r)+\left[\frac{\gamma}{r}+\frac{\delta}{r-1}+\frac{\epsilon_1}{r-b}+\frac{\epsilon_2}{r-a}\right]{F^\prime(r)}+\frac{\alpha\beta r^2+p_1r+p_2}{r(r-1)(r-b)(r-a)}F(r)=0
\end{equation}

We want to compare the equation (12) with equation (13), in that case one can rewrite the equation of (12) in form of Huen equation which is given by,
\begin{equation}
\partial_r^2+\left[\frac{1}{r}+\frac{0}{r-1}+\frac{1}{2(r-r_-)}+\frac{1}{2(r-r_+)}\right]\partial_r - \frac{\ell(\ell+1)r(r-1)}{r(r-1)(r-r_-)(r-r_+)}
\end{equation}

So, by Comparing  the equation (12) and (13) we will have  following relations,
\begin{eqnarray}
\gamma=1,\qquad  \delta=0, \qquad \epsilon_1=\frac{1}{2},\qquad \epsilon_2=\frac{1}{2},\qquad\alpha \beta=-\ell(\ell+1)\nonumber\\
 p_1=\ell(\ell+1),\qquad p_2=0,\qquad     a=r_+, \qquad   b=r_-
\end{eqnarray}
If we take $\alpha=-\ell$ and $\beta=\ell+1$, we will have following condition,
\begin{equation}
\alpha+\beta+1=\gamma+\delta+\epsilon_1+\epsilon_2     \Longrightarrow    2=2
\end{equation}
Now, we are going to show that the corresponding Hamiltonian will be form of $sl(2)$ algebra. As we know, any known second order differential equation can be written  in terms of$ f_1(x)$, $f_2(x)$ and $f_3(x)$ which is given by the following equation
\begin{equation}
\left[f_1(r){\partial_r}^2+f_2(r){\partial_r}+f_3(r)\right]F(r)=0
\end{equation}
which is also satisfied by the following commutation relations and also will be formed of $sl(2)$ closed algebra
\begin{equation}
[P^0,P^+]=A P^+     \hspace{1cm}      [P^0,P^-]=-AP^-    \hspace{1cm}      [P^+,P^-]=FP^0
\end{equation}
If we consider (12) a following.
\begin{eqnarray}
\hat{H^2}&=&\frac{\hbar^2 c^2}{r^3(r-1)}[r(r-1)(r-r_-)(r-r_+){\partial_r}^2+ \nonumber\\&&                                                        ((r-1)(r-r_-)(r-r_+)+\frac{1}{2}r(r-1)(r-r_+)+\nonumber\\&&\frac{1}{2}r(r-1)(r-r_-)){\partial_r}-\ell(\ell+1)r(r-1)]
\end{eqnarray}

We will arrange such equation inform of equation (17), where $f_1(x)$, $f_2(x)$ and $f_3(x)$ as,
\begin{eqnarray}
{f_1(r)}&=&{a_0r^4+a_1r^3+a_2r^2+a_3r+a_4} \nonumber\\
{f_2(r)}&=&{a_5r^3+a_6r^2+a_7r+a_8}\nonumber\\
{f_3(r)}&=&{a_9r^2+a_{10}r+a_{11}}
\end{eqnarray}

We put (20) into equation (17) and compare with (19), one can rewrite the coefficients $a_{i}$($i=0, 1, 2, ...                                                                                11$) as [8].
\begin{eqnarray}
{a_0=1\qquad  a_1=-(r_+ +r_-+1)\qquad a_2=r_+r_-+(r_+ +r_-)\qquad a_3=-r_+r_-  }\nonumber\\
{a_4=0\qquad a_5=2\qquad a_6=-\frac{1}{2}(r_++r_-)\qquad a_7=(1+r_-)(r_++r_-)  }\nonumber\\
{a_8=-r_-r_+ \qquad a_9=-\ell(\ell+1)\qquad a_{10}=\ell(\ell+1)  \qquad a_{11}=0}\nonumber\\
\end{eqnarray}
In order to construct the generalized sl(2) algebra for $\hat{H^2}$, we choose suitable operators $P^+(r)$ , $P^-(r)$ , $P^0(r)$ and $F$ which are given by the following expressions[8]
\begin{eqnarray*}
  a &=& \frac{-3\left[r_+r_-+(r_++r_-)\right]}{\ell} \\
  b &=& \frac{3\left[r_+r_-+(r_++r_-)\right]-2\ell\left[2r_+r_-+2(r_++r_-)+r_-^2\right]}{\ell^2} \\
  c &=& \frac{2\ell^3(\ell+1)[r_+r_-+(r_++r_-)]}{\ell^3}+\nonumber\\&&
 \frac{ \ell(\ell+2)[r_-(r_--r_+)-(r_++r_-)]-3[r_+r_-+(r_++r_-)]}{\ell^3} \\
  d &=& \frac{r_-^2\ell^4+(3-2\ell^3)\left[r_+r_-+(r_++r_-)\right]-\ell(\ell+2)[r_-(r_--r_+)-(r_++r_-)]}{\ell^3}
\end{eqnarray*}

\begin{eqnarray}
P^+&=&r^4{\partial_r}^2+2r^3\partial_r-\ell(\ell+1)r^2\nonumber\\
P^-&=& 2 r\left[r_+r_-+(r_++r_-)\right]{\partial_r}^2+2 \left[(r_++r_-)(1+r_-)\right]\partial_r\nonumber\\
P^0 &=& \ell r\partial_r+(\ell+1)\nonumber\\
F&=& 4ar^3{\partial_r}^2+4br^2\partial_r+4cr\nonumber\\
D&=& d\nonumber\\
\end{eqnarray}
\begin{equation}
[P^0,P^+]=2\ell P^+     \hspace{1cm}      [P^0,P^-]=-2\ell P^-    \hspace{1cm}      [P^+,P^-]=FP^0+D(\ell+1)
\end{equation}
 We note that some symmetry here lead us to investigate   the corresponding black hole from algebra point of view. We see the $R-N$ background  with Klein -Godon particle has a some symmetry and satisfied by generalized $sl(2)$ algebra. And the above commutation relation complectly agree with equation (18) by some generalization.

\section{Energy spectrum of  particle near the proximity  Reissner–Nordström  black hole}
Now,  we apply equation (3) on the $\Psi=Y_\ell^m(\theta,\varphi) F(r)$ and also consider equation (7), so we have following,
\begin{equation}
{\hat{H}}^2\Psi=E^2\Psi
\end{equation}
and
\begin{equation}
\sqrt{-f(r)}\partial_r\left(r^2\sqrt{-f(r)}\partial_rF(r)\right)+\left(\ell(\ell+1)+\lambda r^2\right) F(r)=0
\end{equation}
where $\lambda$ is,
\begin{equation}
\lambda=\frac{E^2-{E_0}^2}{\hbar^2c^2}
\end{equation}
Now, we put  equation (10) into  equation (25) one can rewrite the following relation,
\begin{eqnarray}
F^{\prime\prime}(r)+\left[\frac{1}{r}+\frac{1}{2(r-r_-)}+\frac{1}{2(r-r_+)}\right]F^{\prime}(r)+\nonumber\\ \left[\frac{\frac{\lambda r_-^2+\ell(\ell+1)}{(r_+-r_-)}}{(r-r_-)}-\frac{\lambda r +\frac{\lambda r_+r_-+\ell(\ell+1)}{r_+-r_-}}{(r-r_+)}\right]F(r)=0
\end{eqnarray}

In order to solve this equation one can assume  $r\rightarrow r_+$.  That  limit is  near the proximity black hole.  So, in this limit we rewrite  the  equation (27) as,
\begin{equation}
{F}^{\prime\prime}(r)+\frac{1}{2(r-r_+)}F^{\prime}(r)-\frac{\lambda r +\frac{\lambda r_+r_-+\ell(\ell+1)}{r_+-r_-}}{(r-r_+)}F(r)=0
\end{equation}
By  using  the variable $ x=r - r_+$ the above equation will be as,
\begin{equation}
x{F^{\prime\prime}}(r)+\frac{1}{2}{F^{\prime}}(r)-\left[\lambda x + \frac{\lambda r_+^2+\ell(\ell+1)}{r_+-r_-}\right]F(r)=0
\end{equation}
For obtaining the energy spectrum we need to write the corresponding equation in form of known special function. So, we  choose some change of variable as $F(x)=e^{-\sqrt{\lambda }x} x^\frac{1}{2} G(x)$. In that case the above equation will be as,
\begin{equation}
x{G^{\prime\prime}}(r)+\left[\frac{3}{2}-2\sqrt{\lambda}x\right]{G^{\prime}}(r)-\left[\frac{3}{2}\sqrt{\lambda}+\frac{\lambda r_+^2+\ell(\ell+1)}{r_+-r_-}\right]G=0
\end{equation}
Such equation lead us to introduce  following associated
Laguerre differential equation [9,10],
\begin{equation}
x{L^{\prime\prime}}_{n,m}^{(\alpha,\beta)}(x)+(1+\alpha-\beta x){L^{\prime}}_{n,m}^{(\alpha,\beta)}(x)+\left[(n-\frac{m}{2})\beta-\frac{m}{2}(\alpha+\frac{m}{2})\frac{1}{x}\right]{L}_{n,m}^{(\alpha,\beta)}(x)=0
\end{equation}
If we want to compare our corresponding equation with associated Laguerre we have to consider  $m=0$. So, in that case the associated Laguerre equation will be following form,
\begin{equation}
x{L^{\prime\prime}}_{n,0}^{(\alpha,\beta)}(x)+(1+\alpha-\beta x){L^\prime}_{n,0}^{(\alpha,\beta)}(x)+n\beta \hspace {0.2cm}{L}_{n,0}^{(\alpha,\beta)}(x)=0
\end{equation}

Now we compare equation (30) from black hole side  with associated Laguerre equation (32), one can obtain following results,
\begin{equation}
\alpha=1   \hspace{1cm}   \beta= 2\sqrt{\lambda} \hspace{1cm} n\beta=-\frac{3}{2}\sqrt{\lambda}-\frac{\lambda r_+^2+\ell(\ell+1)}{r_+-r_-}
\end{equation}

Here the parameter $\lambda$ play important role, because it helps us to find the energy spectrum with suitable approach. By using $\beta$ in (33),  take $\lambda$ as parameter  and equation (26) one can obtain the following condition,
\begin{equation}
(2n+\frac{3}{2})^2 \geq \frac{4\ell(\ell+1)r_+^2}{(r_+-r_-)^2}
\end{equation}
We take advantage from this comparatione  and achieve  the following value for the parameter,
 $\lambda$
\begin{equation}
\lambda=\frac{1}{4}\left(\frac{r_+-r_-}{r_+^2}\right)^2\left(2n+\frac{3}{2}\right)^2
\end{equation}
As a mentioned before the energy spectrum complectly depend to $\lambda$ as equation (26)  which is given by,
\begin{equation}
E^2=\frac{1}{4}\left(\frac{r_+-r_-}{r_+^2}\right)^2\left(2n+\frac{3}{2}\right)^2+E_0^2
\end{equation}
By using the free particle condition  as $ E_0=0$ , the  corresponding energy spectrum is obtained by following equation,
\begin{equation}
 E=\frac{1}{2} \hbar c \left(\frac{r_+-r_-}{r_+^2}\right)\left(2n+\frac{3}{2}\right)
\end{equation}

By comparing the energy spectrum of particle moving near Reissner - Nordstrm black hole  with  energy of harmonic oscillator in three dimensions, we have following results,
\begin{equation}
\frac{c}{2} \frac{(r_+-r_-)}{r_+^2}\rightarrow{\omega},\qquad  2n\rightarrow  n
\end{equation}
where  $\omega$ and $n$ are two parameters of harmonic oscillator.
We note here the equation (37) similar to energy spectrum of  harmonic oscillator in three dimensions.
In that case for energy spectrum of harmonic oscillator in three dimensions we have  $n=0,1,2,....$ but for the energy spectrum of particle moving in near Reissner - Nordstrm black hole we have $n=0,2,4,......$

By comparing the energy spectrum of particle moving near Reissner - Nordstrm black hole  with  energy of harmonic oscillator in one dimensions, we have following results,
\begin{equation}
\frac{c}{2} \frac{(r_+-r_-)}{r_+^2}\rightarrow\omega, \qquad 2n+1 \rightarrow n
 \end{equation}
We note again here, the equation (37) will be similar to energy spectrum of the harmonic oscillator in one dimensions.
In that case for energy spectrum of harmonic oscillator in one dimensions we have $n=0,1,2,....$ but for the energy spectrum of particle moving in near Reissner - Nordstrm
black hole we have $ n=1,3,5,......$
Finally,  here one can say that the energy spectrum of the Klein Gordon particle near the Reissner - Nordstrm black hole is the harmonic oscillator energy in three dimensions with even integer or as harmonic oscillator in one dimension with odd integer.\\
Generally we note that when the mass of the black hole increases, the particle frequency increases  But when the electric charge of a black hole increases, its frequency decreases.
\section{Thermal properties of   Reissner - Nordström black hole with energy spectrum}
 Now we are going to use the energy  spectrum of the particle in Reissner - Nordström black hole and  study the thermal properties of system. So, first we obtain entropy and temperature and compare such temperature with Hawking radiation. In order to do such process we first consider the von Neumann entropy which is given by,
\begin{equation}
S=-k_B\sum( p_n\ln p_n)
\end{equation}

Here, we have  probability density operator $p_i$ and the notation $\sum$
is the trace performed in ordinary Fock space representation.
Also, the state function  $<E>$ is the expectation value of
$E_n$ given by,
\begin{equation}
<E>=\sum (p_nE_n)
\end{equation}

The maximum  $S$  can obtained by holding
this  fixed average  and some partition function, of course which is specified by following relation,
value we specified, that is
\begin{equation}
S=k_B(\ln Z+\beta<E>)
\end{equation}

On the other hand,  the partition function for the canonical ensemble is given by
\begin{equation}
Z=\frac{Q^N}{N!}
\end{equation}
\begin{equation}
Q=\sum_{n=0}^{\infty}e^{-\beta E_n}
\end{equation}
\begin{equation}
 \beta = \frac{1}{k_B T}
\end{equation}
Here, we use the above obtained  energy spectrum from corresponding black hole and write the associated partition function and study the thermal properties of system. In that case the
partition function is given by,
\begin{equation}
Q=\sum_{n=0}^{\infty}e^{-\beta E_n}=\sum_{n=0}^{\infty}e^{-\beta  \hbar \omega \left(2n+\frac{3}{2}\right)}
\end{equation}
One can rewrite the above equation as,
\begin{equation}
Q=\frac{e^{-\frac{3}{2}\beta \hbar \omega}}{1-e^{-2\beta \hbar \omega}}
\end{equation}
 We note that the average energy will be following,
 \begin{equation}
 <E>=-\frac{\partial \ln Z}{\partial \beta}
\end{equation}

We use Equation (43) (44) and (48) we will arrive at  following  average value of energy,
\begin{equation}
 <E>=\frac{3}{2} N \hbar \omega + N \frac{2\hbar \omega}{e^{2\beta \hbar \omega}-1}
\end{equation}

Here, one can say that the specific heat capacity can be obtained directly by following formula,
\begin{equation}
  C_V=\frac{\partial < E>}{\partial T}
  \end{equation}

So, the equation (48) into (49) lead us to obtain  specific heat capacity as,
\begin{equation}
  C_V=\frac{4N \hbar ^2 \omega^2}{k_B T^2} \frac{e^{2\beta \hbar \omega}}{(e^{2\beta \hbar \omega}-1)^2}
  \end{equation}
  Putting $T=\frac{1}{k_B \beta}$  into equation (50) and simplifying the corresponding specific heat capacity will be as,
\begin{equation}
   C_V= N k_B \left(\frac{\beta \hbar \omega}{\sinh(\beta \hbar \omega)}\right)^2
\end{equation}
If we apply some constraint as $\beta \hbar \omega\ll 1$ to above equation, we will arrive at following specific heat capacity,
\begin{equation}
  C_V=N k_B
\end{equation}
In order to calculate the entropy of system we  use Equation (43) (46) and (49), so we have following,
\begin{equation}
S=\frac{2Nk_B \beta \hbar \omega}{e^{2\beta \hbar \omega}-1}-Nk_B \ln(1-e^{-2\beta \hbar \omega})-k_B\ln N!
\end{equation}
Here, generally we will try to compare the temperature of black with Hawking temperature. And we show the relation between temperatures in two point of view, in that case we see that such relation is simplified by the suitable physical condition. All the above information lead us to calculate first surface gravity  in the Reissner–Nordström black hole  which is given by,
  \begin{equation}
k_+=\frac{\hbar c}{k_B} \frac{r_+-r_-}{2r_+^2}
\end{equation}

where ${\omega}=c \hspace {0.1cm}\frac{r_+-r_-}{2r_+^2}$ . So such comparing help us to obtain following result,
 \begin{equation}
 \omega=\frac{k_B}{\hbar} k_+
\end{equation}

On the other hand the Hawking temperature for the Reissner–Nordström black hole will be as,
 \begin{equation}
 T_H=\frac{k_+}{2\pi}
\end{equation}

$\omega$ in terms of Hawking temperature will be as,
\begin{equation}
\omega=\frac{2\pi k_B}{\hbar} \hspace {0.2cm} T_H
\end{equation}
The equation (58) shows the relation between free particle frequency of the Klein–Gordon near the Reissner–Nordström  and  Hawking radiation of the black hole.
In order again to write the corresponding temperature in terms of entropy of black hole we first consider the following Stirling's approximation which is given by,
\begin{equation}
\ln N!=N \ln N-N
\end{equation}

By using condition $\beta \hbar \omega \ll 1$ and equation (58) into (53), the entropy is obtained following relation,
\begin{equation}
S=-Nk_B\ln(2 N\beta \hbar \omega)+ 2Nk_B
\end{equation}
where $\beta = \frac{1}{k_B T}$ and  $T \gg 2\pi  T_H.$

The above equation help us to obtain the temperature in terms of entropy which is given by,
\begin{equation}
T=\frac{2N\hbar \omega}{k_B} e^{\frac{S}{N k_B}-2}
\end{equation}
We take $\omega$ in equations (58) one can rewrite $T$ as,
 \begin{equation}
  T=4\pi N T_H  e^{\frac{S}{N k_B}-2}
  \end{equation}
If temperature does not function of entropy we will have following equation,
\begin{equation}
  S=2 N k_B
  \end{equation}
In that case the corresponding temperature is obtained by 
  \begin{equation}
  T=4 \pi N T_H
  \end{equation}
As we know the corresponding
expression proportional to the area of the black hole is
called Bekenstein-Hawking (BH) entropy [11], [12] and [13],  which is defined
by following relation,
\begin{equation}
 S=k_B\frac{A}{4 \ell_p^2}
  \end{equation}

The area of the horizon is obtained by comparing the equation (62) and (64),
\begin{equation}
 A=8 N \ell_p^2
  \end{equation}

  One can think about the boundary of $N- R$ black hole as a storage device for information. Assuming
that the holographic principle holds, the maximal storage space, or total number of
bits, is proportional to the area A. In fact, in an theory of emergent space this how
area may be defined: each fundamental bit occupies by definition one unit cell ${{l_p}^2}.$
Let us denote the number of used bits by $N_0$. It is natural to assume that this
number will be proportional to the area. So we write [14]

The area of the horizon is obtained by comparing the equation (62) and (64):
\begin{equation}
 N_0=\frac{Ac^3}{G\hbar}      \hspace{1cm}   \Longrightarrow     \hspace{1cm}    N_0=\frac{A}{\ell_p^2}
  \end{equation}

  We use equations (66) and (67) one can obtain $ N_0=8 N$.  It shows that each particle in near $R- N,$  $n$ dimension  has a ${2^{n-1}}$ bits of information. For our case , the boundary has three dimension so, we have 8 bit of information  for each particle.

\section {Conclusion}

In this paper, we introduced  Klein Gordon  particle near Reissner–Nordström  black hole. In order to investigated the symmetry and generalized algebra we compared the corresponding Laplace equation with the generalized Heun functions. Such relation lead us to obtained the generalized $sl(2)$ algebra and some suitable results. In order to obtain the energy spectrum we considered
 $r\rightarrow r_+$ which  is near the proximity black hole. In that case the corresponding equation will be form of  Laguerre differential equation. The information of such polynomial helped us to arrange the energy spectrum as form of three dimensional harmonic oscillator energy. Here, also we took advantage from harmonic oscillator energy  with partition function and investigated the thermal properties of black hole. Finally the corresponding temperature and entropy lead us to studied black hole as bit of information.

\end{document}